\documentclass[letterpaper,11pt]{article}
\pdfoutput=1 

\usepackage{jheppub} 

\usepackage{graphicx}
\usepackage{epstopdf}
\usepackage{amsmath, amssymb}
\allowdisplaybreaks

\usepackage{tikz}
\usepackage{verbatim}
\usepackage{bm}
\usepackage{caption}
\usepackage{subcaption}
\usepackage{float}
\usepackage{cellspace}
\usepackage{longtable}
\usepackage{ytableau}
\ytableausetup{smalltableaux, aligntableaux = center}

\def\vec#1{\bm{#1}}

\newcommand{\be}{\begin{eqnarray}}
\newcommand{\ee}{\end{eqnarray}}

\newcommand{\bn}{\begin{enumerate}}
\newcommand{\en}{\end{enumerate}}

\def\IZ{\mathbb{Z}}

\def\CN{{\cal N}}
\def\CO{{\cal O}}

\def\CS{{\cal S}}
\def\CT{{\cal T}}

\def\CV{{\cal V}}

\def\n{\nu}

\def\G{\Gamma}

\def\half{\frac{1}{2}}

\def\Tr{{\rm Tr}}

\usetikzlibrary{arrows}
\usetikzlibrary{positioning}
\usetikzlibrary{decorations.pathmorphing}
\usetikzlibrary{decorations.markings}
\def\arrowhead{angle 90}
\tikzset{>=\arrowhead}
\pgfarrowsdeclaredouble{<<}{>>}{\arrowhead}{\arrowhead}
\pgfarrowsdeclaretriple{<<<}{>>>}{\arrowhead}{\arrowhead}
\pgfarrowsdeclaredouble{<<<<}{>>>>}{<<}{>>}
\tikzstyle{W}=[draw, circle, minimum size=2em, scale=1, inner sep=2pt]
\tikzstyle{B}=[draw,circle,fill=black,scale=1]
\tikzstyle{D}= [circle, minimum size=1em]
\tikzstyle{H}=[draw,circle,fill=gray,scale=1]
\tikzstyle{R}=[draw, inner sep=4pt, minimum size=1.8em]
\tikzstyle{every picture}=[scale=1,baseline=(current bounding box.south)]

\title{A Family of Vertex Algebras from Argyres-Douglas Theory}

\author{Heeyeon Kim}
\author{and Jaewon Song}
\affiliation{Department of Physics, Korea Advanced Institute of Science and Technology \\ 291 Daehak-ro, Yuseong-gu, Daejeon 34141, Republic of Korea}
\emailAdd{heeyeon.kim@kaist.ac.kr}
\emailAdd{jaewon.song@kaist.ac.kr}

\abstract{
We find that multiple vertex algebras can arise from a single 4d $\mathcal{N}=2$ superconformal field theory (SCFT).
The connection is given by the BPS monodromy operator $M$, which is a wall-crossing invariant quantity that captures the BPS spectrum on the Coulomb branch. For a class of low-rank Argyres-Douglas theories, we find that the trace of the multiple powers of the monodromy operator $\Tr M^N$ yield modular functions that can be identified with the vacuum characters of certain vertex algebra for each $N$. 
In particular, we realize unitary VOAs of the Deligne-Cvitanovi\'c exceptional series type $(A_2)_1$, $(G_2)_1$, $(D_4)_1$, $(F_4)_1$, $(E_6)_1$ from Argyres-Douglas theories. We also find the modular invariant characters of the `intermediate vertex algebras' $(E_{7\frac{1}{2}})_1$ and $(X_1)_1$.  
Our analysis allows us to construct 3d $\mathcal{N}=2$ gauge theories that flow to $\mathcal{N}=4$ SCFTs in the IR, whose specialized half-index can be identified with these modular invariant characters.
}

\begin{document}
	
\maketitle

\section{Introduction}

Four-dimensional $\CN=2$ supersymmetric field theories have been known to have connections to two-dimensional field theories in multiple ways, where Seiberg-Witten geometry \cite{Seiberg:1994rs, Seiberg:1994aj} plays a vital role. One connection involves two distinctive objects: the spectrum of operators in a conformal field theory and the spectrum of massive particles. It is a surprising connection since a conformal field theory does not have a particle state, and we only measure the critical exponents and correlators of primary operators. 

The connection is given as follows: All $\CN=2$ supersymmetric theories are believed to have a Coulomb branch where low-energy dynamics is governed by $U(1)^r$ abelian gauge theory with massive charged particles. Such a particle spectrum can be determined from the Seiberg-Witten geometry, which exhibits wall-crossing phenomena as one moves in the Coulomb branch of the moduli space. Thanks to the extended supersymmetry, the protected part of the particle spectrum (BPS particle) is remarkably robust and sometimes can be determined precisely. On the other hand, at the origin of the moduli space (or at a conformal point), we do not have massive BPS particles; instead, we have a conformal field theory. The superconformal field theory has a protected sector, called the Schur sector, which can be mapped to a vertex operator algebra (VOA) \cite{Beem:2013sza}. It maps a unitary 4d SCFT $\CT$ to a non-unitary VOA $\CV$. 
Surprisingly, these two data - the spectrum of BPS particles and the spectrum of primary operators at the CFT point - turn out to be related in a precise way. Namely, the Schur index (defined as a limit of the superconformal index \cite{Kinney:2005ej, Gadde:2011ik, Gadde:2011uv}) of the CFT point, which computes the vacuum character of the VOA, can be written as \cite{Cordova:2015nma}
\begin{align}\label{eq:Schur}
    I_S (q) = \chi_{\CV}(q) = (q)_\infty^{2r} \, \Tr \, M(q)^{-1} \ , 
\end{align}
where the operator $M(q)$ is the quantum monodromy operator in the quantum torus algebra \cite{Cecotti:2010fi}, constructed out of the spectrum of the BPS operators at a given point in the Coulomb branch. Here $r$ is the rank (the dimension of the Coulomb branch) of the 4d theory, and $(q)_\infty \equiv \prod_{n=1}^\infty(1-q^n)$ is the q-Pochhammer symbol.\footnote{Important piece of evidence for the identification of \eqref{eq:Schur} comes from the independent computation of the Schur index from the TQFT description of the AD theories \cite{Buican:2015ina, Buican:2015hsa, Buican:2015tda, Song:2015wta}.}

One of the most important features of the monodromy trace \eqref{eq:Schur} is that it is a wall-crossing invariant quantity, and therefore it is independent of the choice of a Coulomb vacuum. A natural generalization of this formula is to consider the trace of other powers of the monodromy operator, which is, by construction, also a wall-crossing invariant. It was already observed in \cite{Cecotti:2010fi, Cecotti:2015lab} that the trace of this operator $\Tr M$ (without inverse) is given by the characters of various 2d vertex operator algebras. 
Higher powers of the monodromy operator, $\Tr M^N$ for $N>1$ are considered in \cite{Cecotti:2015lab} (see also \cite{Iqbal:2012xm}) and it was conjectured that there is a family of associated chiral algebras $\{\CV^{(N)}\}$ for a given 4d $\CN=2$ SCFT, whose vacuum character is computed by the formula $\Tr M^N$. 

In the current paper, we verify this by explicitly computing $\Tr M^N$ for a number of Argyres-Douglas theories, and show that they reproduce a particular class of modular functions.\footnote{Throughout the paper, by a  modular function we mean a function that is modular under a congruence subgroup of $SL(2,\mathbb{Z})$.} In most cases, we find that it coincides with the vacuum character of a rational VOA whose central charge coincides with that of $\CV^{(N)}$, given by
\begin{align}
    c^{(N)}_{2d} = 12 N c_{4d} -2r(N+1) \ .   
\end{align}

This gives rise to the following schematic diagram:
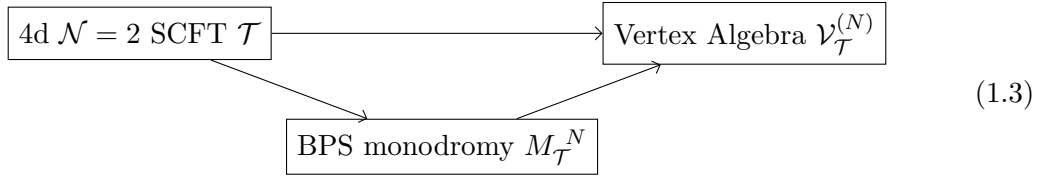
\begin{align}
\begin{split}
      \begin{tikzpicture}
			\node[R] (1) at (0,0) {4d $\CN=2$ SCFT $\mathcal{T}$};
		 	\node[R] (2) at (8,0) {Vertex Algebra $\CV^{(N)}_\CT$};
            \node[R] (3) at (4,-1.5) {BPS monodromy $M_\CT^{~N}$};
			\draw[->] (1)--(2);
            \draw[->] (3)--(2);
            \draw[->] (1)--(3);
  \end{tikzpicture}
\end{split}
\end{align}
This correspondence extends the SCFT/VOA correspondence of \cite{Beem:2013sza} ($N=-1$ in our notation) to a family of vertex algebras labeled by $N$, which also includes \emph{unitary} chiral algebras, from a 4d $\CN=2$ SCFT. 
For example, from the minimal $H_0 = (A_1, A_2)$ Argyres-Douglas theory \cite{Argyres:1995jj}, we obtain a series of vertex algebras ($N=1, 2, 3, 4$): 
\begin{align}
    (A_1, A_2): \quad \mathrm{osp}(1|2)_1, \ (G_2)_1, \  (F_4)_1, \ `` (E_{7\half})_1 " \ , 
\end{align}
with central charges $c=\frac{2}{5}, \frac{14}{5}, \frac{26}{5}, \frac{38}{5}$ respectively. The $N=-1$ case is the previously well-known case of the Virasoro minimal model $M(2,5)$ with $c=-\frac{22}{5}$, and $N=1$ gives the $\mathrm{osp}(1|2)_1$ algebra with $c=\frac{2}{5}$.\footnote{This is sometimes referred to as the conjugate Lee-Yang model or $M(5, 2)_{\textrm{eff}}$. The corresponding two modular functions are realized as the super-characters of two irreducible modules of simple affine VOA $\mathrm{osp}(1|2)$ at level one, whose central charge is $c=2/5$ \cite{creutzig2018representation,Ferrari:2023fez, Creutzig:2024ljv}.} We notice that the VOAs in the above list are the ones that appear in the classification of RCFTs with two modules by Mathur-Mukhi-Sen (MMS) \cite{Mathur:1988na, Mathur:1988gt}. 
Remarkably, we also find the modular invariant characters\footnote{Here, we use the standard terminology in math literature. By modular invariant characters, we mean that they form a basis of the modular invariant subspace in the space of characters.}
of mysterious `intermediate' vertex algebra $(E_{7\half})_1$ \cite{kawasetsu2014intermediate} from the minimal Argyres-Douglas theory! The $E_{7\half}$ algebra appears as the missing hole in the Deligne-Cvitatoni\'c exceptional series \cite{deligne1996, Cvitanovic2008}
\begin{align}
    A_1 \subset A_2 \subset G_2 \subset D_4 \subset F_4 \subset E_6 \subset E_7 \subset ``E_{7\half}" \subset E_8 \ , 
\end{align}
between $E_7$ and $E_8$ whose construction was given in \cite{LANDSBERG2006143}. The MMS classification precisely yields the affine Lie algebras of the above type with level 1 (except for the $(E_8)_1$, which has only one module). 

In addition, by considering $(A_1, A_3)$, $(A_1, A_4)$ and $(A_1, D_4)$ theories \cite{Argyres:1995xn, Eguchi:1996vu, Eguchi:1996ds}, we obtain:
\begin{align}
\begin{split}
    (A_1, A_3): &\quad (A_2)_1, \ (E_6)_1 \\
    (A_1, A_4): &\quad \mathrm{osp}(1|4)_1, \ ``(X_1)_1" \\
    (A_1, D_4): &\quad (D_4)_1 
\end{split}
\end{align}
This means that we find all the MMS RCFTs from the simplest examples of Argyres-Douglas theories, except for the $(A_1)_1$ and $(E_7)_1$. We also find characters of another intermediate vertex algebra, called $(X_1)_1$ with $c=\frac{52}{7}$.\footnote{Some of these entries above, namely $(A_2)_1, (D_4)_1$, are also found in \cite{Buican:2017rya, Buican:2019huq} which gives another piece of evidence for our correspondence.}

The physical meaning of the higher powers of monodromy operators can be most easily understood from the twisted circle compactification \cite{Cecotti:2010fi,Dedushenko:2023cvd, Gaiotto:2024ioj, ArabiArdehali:2024ysy, ArabiArdehali:2024vli}.\footnote{Precursors to this correspondence was given in \cite{Fredrickson:2017yka, Dedushenko:2018bpp}.} Consider the holomorphic topological twist of a 4d $\CN=2$ SCFT on $C_q\times \mathbb{C}$, where $C_q$ is a topologically twisted Melvin cigar. Compactification along the cigar circle involves a uniform $2\pi$ rotation of the $U(1)_r$ symmetry, which induces a Janus-like configuration along a closed loop in the Coulomb branch effective theory. The 3d BPS particles are then trapped at various loci on the loop according to their central charges, which gives rise to an effective 3d $\CN=2$ Chern-Simons matter theory description \cite{Gaiotto:2024ioj}. These theories are expected to flow to a superconformal theory with $\CN=4$ supersymmetry enhancement, which often admits a boundary condition that supports a rational VOA upon a topological twist \cite{Gang:2018huc, Gang:2021hrd, Gang:2023rei, Gang:2024loa}. This configuration is most interesting for non-Lagrangian SCFTs, where the Coulomb branch operators have fractional $U(1)_r$ charges, as in the Argyres-Douglas theories.

Furthermore, the BPS monodromy trace we study yields the characters in the so-called `fermionic formulae' or the `Nahm sum' form. Such a form naturally appears as a half-index \cite{Gadde:2013wq} of three-dimensional $\CN=2$ supersymmetric gauge theories with suitable boundary conditions \cite{Dimofte:2017tpi}, from which we can construct the corresponding 3d gauge theory \cite{Gaiotto:2024ioj}. 

The higher powers of the monodromy traces can be obtained by considering a twisted compactification along a closed path that winds around the Janus loop multiple times. It is natural to expect that there exists a positive integer $n$ for each theory where the $n$-th wrapping of the Janus loop trivializes the $U(1)_r$ twisting. For Argyres-Douglas theories, this number coincides with the least common multiple of the denominators of the R-charges of the Coulomb branch operators.

The multi-wrapping Janus-loop configurations give rise to a family of 3d $\CN=2$ Chern-Simons matter theories which are expected to flow to either an $\CN=4$ SCFT or a unitary TFT in the infrared. Alternatively, one can obtain such a description from the twisted reduction of a 4d $\CN=1$ Lagrangian description which flows to an $\CN=2$ Argyres-Douglas theory in the infrared \cite{Maruyoshi:2016tqk, Maruyoshi:2016aim, Agarwal:2016pjo}, and considering the ``higher sheets" of the partition function computation\footnote{The `second sheet' of the superconformal index was considered in \cite{Kim:2019yrz, Cabo-Bizet:2019osg, Cassani:2021fyv} to capture the asymptotic density of states in 4d SCFTs.}, as discussed in \cite{Dedushenko:2018bpp,ArabiArdehali:2024ysy, ArabiArdehali:2024vli}.

We can summarize the connections in the following diagram: 
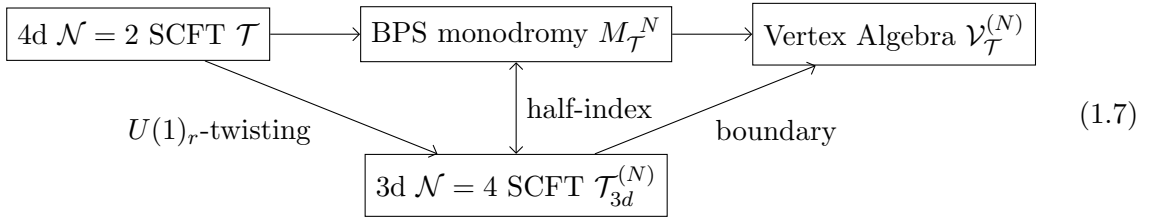
\begin{align}
\begin{split}
      \begin{tikzpicture}
			\node[R] (1) at (0,0) {4d $\CN=2$ SCFT $\mathcal{T}$};
            \node[R] (2) at (5,0) {BPS monodromy $M_\CT^{~N}$};
            \node[R] (3) at (10, 0) {Vertex Algebra $\CV_\CT^{(N)}$};
            \node[R] (4) at (5, -2) {3d $\CN=4$ SCFT $\mathcal{T}_{3d}^{(N)}$};
			\draw[->] (1)--(2);
            \draw[->] (2)--(3);
            \draw[<->] (2)--node[anchor=west] {half-index}(4);
            \draw[->] (4)--node[anchor=north west] {boundary}(3);
            \draw[->] (1)--node[anchor=north east] {$U(1)_r$-twisting}(4);
  \end{tikzpicture}
\end{split}
\end{align}
In this paper, we establish and explore this connection for a small sample of Argyres-Douglas theories.

The organization of this paper is as follows: In section \ref{sec:TrM}, we spell out the details of the correspondence between monodromy trace and vertex algebras. In particular, we perform explicit computations for a series of Argyres-Douglas theories of low rank. In section \ref{sec:3dSCFT}, we focus on two particular cases that give rise to the modular invariant characters of intermediate vertex algebras $(E_{7\half})_1$ and $(X_1)_1$ and construct the $\CN=2$ gauge theories that are expected to flow to 3d $\CN=4$ SCFT in the IR, whose specialized half-index coincides with these modular functions.

\section{Vertex Algebras from Monodromy Traces} \label{sec:TrM}

In this section, we compute the monodromy traces for a number of Argyres-Douglas theories and match them with the characters of vertex algebras. 
Let $\Gamma$ be the electromagnetic charge lattice on the Coulomb branch with an anti-symmetric pairing $\langle ~, ~\rangle$, which contains the flavor lattice $\Gamma_f$. The monodromy operator is defined as
\begin{align}
    M = \prod^{\curvearrowleft}_\gamma \Psi_q (X_\gamma) \ , 
\end{align}
where the product is over all BPS states of charge $\gamma \in \Gamma$ at a given chamber ordered according to the phase of central charges. \footnote{Note that the ordering is reversed compared to the definition of $\CO(q)$ in \cite{Cordova:2015nma}, which is $M^{-1}$ in our convention.} The function $\Psi_q (X)$ is defined as
\begin{align}
    \Psi_q (X) = \prod_{n \ge 0} (1 +q^{\half+n} X) = \sum_{k\ge 0} \frac{q^{\frac{k^2}{2}}}{(q)_k} X^k \ , 
\end{align}
which can be thought of as a partition function of a BPS particle. Here we used the $q$-Pochhammer symbol
\begin{align}
    (z;q)_n \equiv \prod_{i=0}^{n-1} (1-z q^{i}) \ , \quad (q)_n \equiv (q; q)_n \ , \quad (z; q) \equiv (z;q)_\infty \ . 
\end{align}
The monodromy operator $M(q)$ is valued in the quantum torus algebra generated by non-commuting variables $X_\gamma$'s satisfying
\begin{align}
    X_{\gamma_1} X_{\gamma_2} = q^{\half \langle \gamma_1, \gamma_2 \rangle} X_{\gamma_1 + \gamma_2} = q^{\langle \gamma_1, \gamma_2 \rangle} X_{\gamma_2} X_{\gamma_1} \ . 
\end{align}
We will often use the following identities:
\begin{align} \label{eq:ids}
\begin{split}
     &  \Psi_q (X_{\gamma_2}) \Psi_q (X_{\gamma_1}) = \Psi_q (X_{\gamma_1}) \Psi_q (X_{\gamma_1 + \gamma_2}) \Psi_q (X_{\gamma_2})\ ,\quad \text{if } \langle \gamma_1,\gamma_2\rangle =1 \\
    &\Theta(X_{\gamma_2}) \Psi_q (X_{\gamma_1}) = \Psi_q (X_{\gamma_1 + \gamma_2}) \Theta(X_{\gamma_2})\\
    & \Theta(X_\gamma) = \Psi_q (X_{\gamma}) \Psi_q (X_{-\gamma}) = \frac{1}{(q)_\infty} \sum_{n \in \IZ} q^{\frac{n^2}{2}} (X_{\gamma})^n
\end{split}
\end{align}
These identities will turn out to be useful later. 

The trace in the quantum torus algebra can be evaluated by imposing
\begin{align}
\Tr X_\gamma = \left\{\begin{array}{cc} 
     X_{\gamma_f}\ , &\quad\quad  \text{if~} \gamma =\gamma_f\in \G_f \\
    0\ ,& \text{otherwise}
    \end{array}\right. \ .
\end{align}
Since the elements in $\Gamma_f$ are central, we can replace $X_{\gamma_f}$ by a complex number
$x_{f} = \prod_i x_i^{f_i}$, 
where $f_i$ are flavor charges in some basis of $\Gamma_f$.
This means that we impose the Gauss law constraint, taking contributions only from the gauge-invariant, neutral states. We impose a non-vanishing trace only if the corresponding BPS particle has neither electric nor magnetic charge. 

Now let us consider the trace of a higher power of the monodromy operator
\begin{align}\label{higher monodromy prefactor}
    I_N (q) \equiv (q)_\infty^{2r} \, \Tr M^N \ , 
\end{align}
where $r$ is the rank (complex dimension of the Coulomb branch) of the underlying $\CN=2$ theory and $N$ is an integer. When this expression is convergent, we expect that it can be identified with the vacuum character of a vertex  algebra $\mathcal{V}^{(N)}$. 

We conjecture that the central charge $c_{2d}$ of $\mathcal{V}^{(N)}$ is given as 
\begin{align} \label{eq:c2d}
    c^{(N)}_{2d} = 12N c_{4d} - 2r(N+1) = N(12c_{4d} - 2r) - 2r \ , 
\end{align}
where $c_{4d}$ is one of the central charges of the 4d $\CN=2$ SCFT and $r$ being the rank of the 4d theory. One way to see this relation is by invoking the SCFT/VOA correspondence of \cite{Beem:2013sza}, which gives rise to the 2d central charge as $c_{2d} = -12 c_{4d}$. This case corresponds to $N=-1$ in our setup, where the character of the VOA is given by the Schur index of the 4d SCFT. \cite{Cordova:2015nma} 

The higher powers of $M$ has been considered in \cite{Cecotti:2015lab} but with a different prefactor $(q)_\infty^{-2rN}$, giving the effective central charge $c_{2d}^{\textrm{CSVY}} = 12N c_{4d}$. Taking the difference of the prefactor into account, we obtain the central charge \eqref{eq:c2d}. Notice that the prefactor we choose in \eqref{higher monodromy prefactor} is more natural from the point of view of the half-index of the 3d theory.
One can also directly compute the effective central charge by examining the asymptotic behavior of the trace as $q \to 1$. This gives a conjecture for the central charge for the $(G, G')$ Argyres-Douglas theory to be \cite{Cecotti:2015lab}
\begin{align} \label{eq:cAD}
    c_{2d}^{(N)} = \frac{N r_G r_{G'} h_G h_{G'}}{h_G + h_{G'}} -2r \ ,  
\end{align} 
where $r_G$ and $h_{G}$ are the ranks and the dual coxeter numbers of $G$ respectively and $r$ is the rank of the 4d SCFT. 

In the remainder of this section, we test our proposals for a set of Argyres-Douglas theories of low rank. 

\subsection{$(A_1, A_2)$ theory}

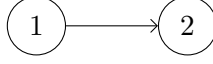
\begin{figure}
  \centering
  \begin{tikzpicture}
			\node[W] (1) at (0,0){1};
		 	\node[W] (2) at (2,0) {2};
			\draw[->] (1)--(2);
  \end{tikzpicture}
  \caption{BPS quiver for the $(A_1, A_2)$ theory}
  \label{fig:A1A2quiver}
\end{figure}

The BPS quiver for the $(A_1, A_2)$ theory is given as in figure \ref{fig:A1A2quiver}.
The monodromy operator \cite{Cecotti:2010fi} for the $(A_1, A_2)$ theory is given as
\begin{align}
    M = \Psi_q(X_{\gamma_2}) \Psi_q (X_{\gamma_1}) \Psi_q(X_{-\gamma_2}) \Psi_q (X_{-\gamma_1}) \ , 
\end{align}
and it can be rewritten as follows:
\begin{align} \label{eq:A1A2Mrel}
\begin{split}
M &= \Theta(X_{\gamma_2}) \Psi_q(X_{\gamma_1-\gamma_2}) \Theta(X_{\gamma_1})\\
&= \Theta(X_{\gamma_2}) \Theta(X_{\gamma_1})\Psi_q(X_{-\gamma_2})\\
&= \Psi_q(X_{\gamma_1}) \Theta(X_{\gamma_2})\Theta(X_{\gamma_1})
\end{split}
\end{align}
In this section, we evaluate the trace of (powers of) the monodromy operator and match it with characters of certain vertex operator algebras. From \eqref{eq:c2d} and \eqref{eq:cAD}, we expect the central charge for the corresponding VOA to be given as
\begin{align}
    c^{(N)}_{2d} = \frac{12N}{5} - 2 \ . 
\end{align}
This gives the central charges for $N=1, 2, 3, 4$ to be 
\begin{align}
     c_{2d} = \frac{2}{5}, \frac{14}{5}, \frac{26}{5}, \frac{38}{5}. 
\end{align}
As we will see, we find that the monodromy traces produce the characters for 
\begin{align}
    \mathrm{osp}(1|2)_1 , (G_2)_1, (F_4)_1, ``(E_{7\half})_1" \ , 
\end{align}
respectively. Of course $N=-1$ case gives us the Lee-Yang minimal model $M(2, 5)$ with $c=-\frac{22}{5}$. The meaning of the last entry in the above list will be explained momentarily.

\subsubsection{$\Tr M$: $\mathrm{osp}(1|2)_1$}
The trace of the monodromy operator can be straightforwardly computed as
\begin{align}
\begin{split}
    \Tr M &= \Tr \, \Psi_q(X_{\gamma_2}) \Psi_q (X_{\gamma_1}) \Psi_q(X_{-\gamma_2}) \Psi_q (X_{-\gamma_1}) \\
     &= \Tr \, \Psi_q(X_{\gamma_1}) \Psi_q (X_{\gamma_1+\gamma_2}) \Psi_q (X_{\gamma_2}) \Psi_q(X_{-\gamma_2}) \Psi_q (X_{-\gamma_1})  \\
     &= \Tr \, \Theta(X_{\gamma_1}) \Psi_q (X_{\gamma_1+\gamma_2}) \Theta(X_{\gamma_2}) \\
     &= \frac{1}{(q)_\infty^2} \sum_{n_1, n_2 \in \IZ}\sum_{k \ge 0} \frac{q^{\frac{n_1^2+n_2^2+k^2}{2}}}{(q)_k} \Tr (X_{n_1 \gamma_1} X_{k(\gamma_1+\gamma_2)} X_{n_2 \gamma_2} ) \\
     &= \frac{1}{(q)_\infty^2} \sum_{n_1, n_2 \in \IZ}\sum_{k \ge 0} \frac{q^{\frac{n_1^2+n_2^2}{2}}}{(q)_k}\Tr (X_{\gamma_1}^{n_1+k} X_{\gamma_2}^{n_2+k} )  \\
     &= \frac{1}{(q)_\infty^2} \sum_{k \ge 0} \frac{q^{k^2}}{(q)_k} \ .
\end{split}
\end{align}
From this, we obtain
\begin{align}
    I_1(q) = (q)_\infty^2 \Tr M = \sum_{n \ge 0} \frac{q^{n^2}}{(q)_n} = \frac{1}{(q; q^5)(q^4; q^5)} \ , 
\end{align}
where the second identity is the Rogers-Ramanujan identity. This gives the character of the non-vacuum module ($h=-\frac{1}{5}$) of the Lee-Yang minimal model (whose central charge is $c=-\frac{22}{5}$, with effective central charge $c_{\textrm{eff}} = c - 24 h_{\textrm{min}} = \frac{2}{5}$) which is identical to the (specialized) vacuum character of the $\mathrm{osp}(1|2)_1$ whose central charge is $c=\frac{2}{5}$. This monodromy trace was already computed in \cite{Cecotti:2010fi, Cecotti:2015lab}.

\subsubsection{$\Tr M^2$: $(G_2)_1$}
Now, let us consider the square of the monodromy operator. Its trace can be computed as
\begin{align}
\begin{split} \label{eq:TrMG2}
    \Tr M^2 &= \Tr \, \Theta(X_{\gamma_1}) \Psi_q (X_{\gamma_1+\gamma_2}) \Theta(X_{\gamma_2}) \Theta(X_{\gamma_1}) \Psi_q (X_{\gamma_1+\gamma_2}) \Theta(X_{\gamma_2}) \\
    &= \frac{1}{(q)_\infty^4}\sum_{\vec{n}, \vec{m} \in \IZ^2} \sum_{k, l \ge 0} \frac{q^{\frac{n_1^2 + n_2^2 + m_1^2 + m_2^2}{2}}}{(q)_k (q)_l} \Tr (X_{\gamma_1}^{n_1+k} X_{\gamma_2}^{n_2+k} X_{\gamma_1}^{m_1+l} X_{\gamma_2}^{m_2+l} ) \\
    &= \frac{1}{(q)_\infty^4}\sum_{\vec{n}, \vec{m} \in \IZ^2} \sum_{k, l \ge 0} \frac{q^{\frac{n_1^2 + n_2^2 + m_1^2 + m_2^2}{2}}}{(q)_k (q)_l} q^{-(n_2+k)(m_1+l)}\Tr (X_{\gamma_1}^{n_1+k+m_1+l}X_{\gamma_2}^{n_2+k +m_2+l} ) \\
    &= \frac{1}{(q)_\infty^4}\sum_{\vec{n} \in \IZ^2} \sum_{k, l \ge 0} \frac{q^{2 k^2+2 k l+2 k n_1 +2 k n_2+l^2+l n_1 + l n_2 + n_1^2+  n_1 n_2 + n_2^2 } }{(q)_k (q)_l} \ . 
\end{split}
\end{align}
Upon multiplying by the prefactor, we obtain
\begin{align}
    I_2(q) = (q)_\infty^2 \Tr M^2 &= 1 + 14 q + 42 q^2 + 140 q^3 + 350 q^4 + 840 q^5 + 1827 q^6 + \ldots \ , 
\end{align}
which agrees with the vacuum character of the $(G_2)_1$ to the leading orders in $q$-series expansion. Notice that $c=\frac{14}{5}$ we predicted above agrees with that of the Sugawara central charge of $(G_2)_1$. We conjecture that the last line of \eqref{eq:TrMG2} gives the vacuum character of the $(G_2)_1$ affine Lie algebra.

\subsubsection{$\Tr M^3$: $({F}_4)_1$}

The trace of the third power of the monodromy operator can be rewritten as
\begin{align}
\begin{split}
    \Tr M^3 &= \Tr \, \Psi_q (X_{\gamma_1}) [ \Theta(X_{\gamma_2}) \Theta(X_{\gamma_1}) \Psi_q (X_{\gamma_1}) ][ \Theta(X_{\gamma_2}) \Theta(X_{\gamma_1}) \Psi_q (X_{\gamma_1}) ] \Theta(X_{\gamma_2}) \Theta(X_{\gamma_1}) \\
    &= \Tr \, \Psi_q (X_{\gamma_1}) \Psi_q (X_{\gamma_1+\gamma_2}) [\Theta(X_{\gamma_2}) \Theta(X_{\gamma_1}) \Psi_q (X_{\gamma_1+\gamma_2})] [\Theta(X_{\gamma_2}) \Theta(X_{\gamma_1})]^2  \\
    &= \Tr \, \Psi_q (X_{\gamma_1}) \Psi_q (X_{\gamma_1+\gamma_2}) \Psi_q (X_{\gamma_2}) [\Theta(X_{\gamma_2}) \Theta(X_{\gamma_1})]^3 \\
    &= \Tr \, \Psi_q (X_{\gamma_2})  \Psi_q (X_{\gamma_1}) \left[\Theta(X_{\gamma_2}) \Theta(X_{\gamma_1}) \right]^3 \ ,
\end{split}
\end{align}
by applying the identities \eqref{eq:ids} and also \eqref{eq:A1A2Mrel}. 
From this, we find
\begin{align}
\begin{split}
    I_3(q) &= (q)_\infty^2 \Tr M^3 = \frac{1}{(q)_\infty^4} \sum_{\vec{n}\in \IZ^6} \sum_{k, l \ge 0} \frac{q^{\frac{n_1^2 + \cdots + n_6^2 + k^2 + l^2}{2}}}{(q)_k (q)_l} \Tr (X_{\gamma_2}^l X_{\gamma_1}^k X_{\gamma_2}^{n_2} X_{\gamma_1}^{n_1} X_{\gamma_2}^{n_4} X_{\gamma_1}^{n_3} X_{\gamma_2}^{n_6} X_{\gamma_1}^{n_5}) \\
    &= \frac{1}{(q)_\infty^4} \sum_{\vec{n}\in \IZ^6} \sum_{k, l \ge 0} \frac{q^{\frac{n_1^2 + \cdots + n_6^2 + k^2 + l^2}{2} + n_1 n_4 + n_3 n_6 + n_6 n_1 -l k }}{(q)_k (q)_l} \Tr (X_{\gamma_1}^{n_1+n_3+n_5+k} X_{\gamma_2}^{n_2+n_4+n_6+l} ) \\
    &= \frac{1}{(q)_\infty^4} \sum_{\vec{n}\in \IZ^4} q^{n_1^2-n_1 n_2+n_1 n_3+n_2^2-n_2 n_3+n_2 n_4+n_3^2-n_3 n_4+n_4^2} \sum_{k, l \ge 0}\frac{q^{k^2-k l+k n_1+k n_3+l^2-l n_1+l n_2-l n_3+l n_4 }}{(q)_k (q)_l}   \\
    &= 1+ 52 q + 377 q^2 + 1976 q^3 + 7852 q^4 + 27404 q^5 + 84981 q^6 + 243230 q^7 + \cdots \ , 
\end{split}
\end{align}
which agrees with the vacuum character of the $(F_4)_1$ to the leading orders in $q$-series expansion. The central charge $c=\frac{26}{5}$ agrees with the Sugawara central charge of $(F_4)_1$. We conjecture that this gives the vacuum character of the $(F_4)_1$ affine Lie algebra. 

\subsubsection{$\Tr M^4$: $({E}_{7 \frac{1}{2}})_1$}

Using the identity \eqref{eq:A1A2Mrel}, we can rewrite $\Tr M^4$ as
\begin{align}
\begin{split}
\Tr M^4 &= \Tr~ \Theta(X_{\gamma_2}) \Theta(X_{\gamma_1}) \Psi_q(X_{-\gamma_2}) \left[\Psi_q(X_{\gamma_1})\Theta(X_{\gamma_2})\Theta(X_{\gamma_1})\right]^3 \\
& = \Tr~ \Theta(X_{\gamma_2}) \Theta(X_{\gamma_1}) \Psi_q(X_{-\gamma_2}) \Psi_q(X_{\gamma_1}) \Psi_q(X_{\gamma_1+\gamma_2})\Psi(X_{\gamma_2}) \left[\Theta(X_{\gamma_2}) \Theta(X_{\gamma_1}) \right]^4 \\
& = \Tr ~ \Psi(X_{\gamma_1}) \left[\Theta(X_{\gamma_2}) \Theta(X_{\gamma_1}) \right]^4 \Theta (X_{\gamma_2})\ .
\end{split}
\end{align}
Expanding the last expression, we find
\begin{align}
\begin{split}
\Tr M^4 = \frac{1}{(q)_\infty^9}\sum_{n\geq 0}\frac{1}{(q)_n}\sum_{\{m_1,\cdots, m_8,k\} \in \mathbb{Z}^9} & q^{\frac12 (n^2 + k^2 + m_1^2 + \cdots m_8^2)}q^{-\frac12\left[m_1(m_2+m_4+m_6+m_8)+m_2m_1\right]}\\
& \cdot q^{-\frac12\left[m_3(m_4+m_6+m_8) + m_4(m_3+m_1) + m_5 (m_6+m_8)\right]}\\
& \cdot q^{-\frac12\left[m_6(m_5+m_3+m_1) + m_7 m_8+ m_8(m_7+m_5+m_3+m_1)\right]}\\
& \cdot \delta_{m_2+m_4+m_6+m_8+n, 0} \cdot \delta_{ m_1+m_3+m_5+m_7+k, 0} \ . 
\end{split}
\end{align}
Therefore
\be
I_4(q) = (q)_\infty^2 \Tr M^4 =  \frac{1}{(q)_\infty^7}\sum_{n\geq 0}\frac{1}{(q)_n}\sum_{{(m_1, \cdots, m_7)} \in \mathbb{Z}^7} q^{\frac12 {\bf m}^tK_8 {\bf m}}\ ,
\ee
where ${\bf m}= (n,m_1,\cdots, m_7)$ and
\be\label{K8 def}
K_8 = \left(\begin{array}{cccccccc}2 & 1 & 1& 1 & 1& 1& 1& 1 \\ 1 & 2 & 0 & 1 & 0 & 1 & 0 & 1\\ 1 & 0 & 2 & 1 & 1 & 1 & 1 & 1 \\ 1 & 1 & 1 & 2 & 0 & 1 & 0 & 1\\ 1 & 0 & 1 & 0 & 2 & 1 & 1 & 1 \\ 1 & 1 & 1 & 1 & 1& 2 & 0 & 1\\1 & 0 & 1 & 0 & 1 & 0 & 2 & 1\\1 & 1 & 1 & 1 & 1 & 1 & 1 & 2\end{array}\right)
\ee
 Note that the matrix $K_8$ defines a rank-8 unimodular even lattice, which must be isomorphic to the $E_8$ lattice. In fact, we can construct a $SL(8,\mathbb{Z})$ matrix $U$ such that $K_8=U C(E_8) U^T$, where
\be
U = \left(\begin{array}{cccccccc} 1 & 0 & 0 & 0 & 0 & 0 & 0 & 0\\ 0 & -1 & 0 & 0 & 0 & 0 & 0 & 0\\  0 & -1 & -2 & -2 & -2  & -1 & 0 & -1 \\ 0 & -1 & -1 & 0 & 0 & 0 & 0 & 0\\ 0 & -1 & -2 & -3 & -4 & -2 & -1 & -2\\ 0 & -1 & -1 & -1 & -2 & -1 & 0 & -1\\ 0 & -1 & -2 & -3 & -3 & -2 & -1 & -2\\ 0 & -1 & -1 & -1 & -1 & 0 & 0 & -1\end{array}\right) \ . 
\ee
and $C(E_8)$ is the standard Cartan matrix of $E_8$:
\be\label{CK8 def}
C(E_8) = \left(\begin{array}{cccccccc}2 & -1 & 0 & 0 & 0 & 0 & 0&0 \\ -1 & 2 & -1 & 0 & 0 & 0 & 0 & 0\\0 & -1 & 2 & -1 & 0 & 0 & 0 & 0 \\ 0 & 0 & -1 & 2 & -1 & 0 & 0 & 0 \\0 & 0 & 0 & -1 & 2 & -1 & 0 & -1\\ 0 & 0 & 0 & 0 & -1 & 2 & -1 & 0\\0 & 0 & 0 & 0 & 0 & -1 & 2 & 0\\ 0 & 0 & 0 & 0 & -1 & 0 & 0 & 2 \end{array}\right)\ .
\ee
From this, we conclude that the change of basis from \eqref{K8 def} to \eqref{CK8 def} leaves the first basis vector $\alpha_1$ invariant. This leads to the identity \footnote{We notice that the $q$-series in the first line converges much quicker compared to the one in the second line.} 
\begin{align}
\begin{split} \label{nahm E7.5}
I_4(q) = (q)_\infty^2 \Tr M^4 &=  \frac{1}{(q)_\infty^7}\sum_{n\geq 0}\frac{1}{(q)_n}\sum_{(m_1, \cdots, m_7) \in \mathbb{Z}^7} q^{\frac12 {\bf m}^tK_8 {\bf m}} \\
&= \frac{1}{(q)_\infty^7}\sum_{n_1\geq 0}\sum_{\{n_2,\cdots, n_8\}\in \mathbb{Z}^7} \frac{q^{\frac12 {\bf n}^t C(E_8) {\bf n}}}{(q)_{n_1}}\ ,
\end{split}
\end{align}
where ${\bf n} = (n_1,\cdots, n_8)$. The latter expression is nothing but one of the modular invariant characters of the intermediate vertex algebra $(E_{7\frac12})_1$ \cite{kawasetsu2014intermediate}, which has a $q$-expansion
\begin{align} \label{E7.5 character 11}
I_4 = (q)_\infty^2\Tr~ M^4 = 1+ 190q + 2831q^2 + 22306 q^3 + \CO(q^4) \ . 
\end{align}
This is one of the solutions to the second-order modular differential equations that appear in the MMS classification \cite{Mathur:1988gt} with $c=\frac{38}{5}$. This value formally agrees with the Sugawara central charge formula of the affine Lie algebra if we set $h^\vee = 24$ and $\textrm{dim}(E_{7\half}) = 190$. 
Combined with another character whose weight is $h=\frac{4}{5}$, we can construct a modular-invariant partition function with the putative chiral algebra. However, it was pointed out that \cite{Mathur:1988na} this leads to a negative fusion coefficient, so it cannot give a consistent RCFT. 
It is not known to us whether it is possible to have a rational VOA with a vacuum character given above. 
However, it is known that these two modular functions are realized as the characters of two irreducible modules of a rational and $C_2$-cofinite W-algebra $W_{-5}(E_8,f_\theta)$ in the \emph{twisted sector} \cite{kawasetsu2018algebras}.\footnote{We thank T. Arakawa for pointing out this to us.} See also \cite{Lee:2023owa, Sun:2024mfz}.

\subsection{$(A_1, A_3)$ theory}

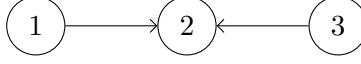
\begin{figure}
  \centering
  \begin{tikzpicture}
			\node[W] (1) at (0,0){1};
		 	\node[W] (2) at (2,0) {2};
            \node[W] (3) at (4,0) {3};
			\draw[->] (1)--(2);
            \draw[<-] (2)--(3);
  \end{tikzpicture}
  \caption{\label{fig:A1A3quiver}
  BPS quiver for the $(A_1, A_3)$ theory in the canonical chamber}
\end{figure}

Let us now consider the simplest example with a flavor symmetry. The BPS quiver in the sink/source chamber for the $(A_1, A_3)=(A_1, D_3)$ theory is given as in figure \ref{fig:A1A3quiver}. 
In this chamber, the monodromy operator can be written as
\begin{align}
    M = \Psi_q (X_{\gamma_2}) \Psi_q (X_{\gamma_3}) \Psi_q (X_{\gamma_1}) \Psi_q (X_{-\gamma_2}) \Psi_q (X_{-\gamma_3}) \Psi_q (X_{-\gamma_1})  \ . 
\end{align}
This theory has a flavor symmetry, and we can choose the flavor lattice vector as 
\begin{align}
    \gamma_f = \gamma_3 - \gamma_1 \ , 
\end{align}
so that 
\begin{align}
    X_{\gamma_3} = X_{\gamma_1} X_{\gamma_f} \ . 
\end{align}
The central charge for the VOA corresponding to $M^N$ is given as
\begin{align}
    c_{2d}^{(N)} = 4N - 2 \ , 
\end{align}
so that we get $c=2$ for $N=1$ and $c=6$ for $N=2$. We argue that they corresponds to $(A_2)_1$ and $(E_6)_1$ respectively.   

\subsubsection{$\Tr M$: $(A_2)_1$}

Upon using the identities we introduced above, we find
\begin{align}
\begin{split}
    \Tr M &= \Tr~ \Psi_q (X_{\gamma_2}) \Psi_q (X_{\gamma_3}) \Psi_q (X_{\gamma_1}) \Psi_q (X_{-\gamma_2}) \Psi_q (X_{-\gamma_3}) \Psi_q (X_{-\gamma_1})  \\
    &= \Tr~\Theta(X_{\gamma_3})\Psi_q(X_{\gamma_2+\gamma_3})\Psi_q(X_{\gamma_1})\Psi_q(X_{\gamma_{1}+\gamma_2})\Theta(X_{\gamma_2})\Psi(X_{-\gamma_1}) \\
    &= \Tr~\Theta(X_{\gamma_1})\Theta(X_{\gamma_2})\Theta(X_{\gamma_3}) \Theta(X_{\gamma_2}) \ . 
\end{split}
\end{align}
From here, we find
\begin{align}
\begin{split} 
    I_1(q) = (q)_\infty^2 \Tr M &=  \frac{1}{(q)_\infty^2} \sum_{\vec{n} \in \IZ^4} q^{\half(n_1^2 + n_2^2 + n_3^2 + n_4^2)} \Tr (X_1^{n_1} X_2^{n_2} X_3^{n_3} X_2^{n_4} ) \\
    &= \frac{1}{(q)_\infty^2} \sum_{\vec{n} \in \IZ^2} q^{\half \vec{n}^T C(A_2) \vec{n}} \\ 
    &= 1 + 8q + 17q^2 + 46 q^3 + 98 q^4 + 198 q^5 + 371 q^6 + \cdots \ , 
\end{split}
\end{align}
where we have used $X_3 = X_1 X_f$ with $\Tr X_f = 1$ and $C(A_2)$ is the Cartan matrix for the Lie algebra $A_2$. This is the theta function for the $A_2$ root lattice and, indeed, is identical to the vacuum character of the affine Lie algebra $(A_2)_1$. 
 
\subsubsection{$\Tr M^2$: $(E_6)_1$}

Upon evaluating the trace of the square of the monodromy operator, we obtain
\begin{align}
\begin{split} 
    \Tr M^2 
&= \Tr~ \Theta(X_{\gamma_1}) \Theta(X_{\gamma_2}) \Theta(X_{\gamma_3}) \Theta(X_{\gamma_2}) \Theta(X_{\gamma_1}) \Theta(X_{\gamma_2}) \Theta(X_{\gamma_3}) \Theta(X_{\gamma_2}) \ . 
\end{split}
\end{align}
Therefore, we get
\begin{align}
\begin{split}
    I_2(q) &= (q)_\infty^2 \Tr M^2 = \frac{1}{(q)_\infty^6} \sum_{\vec{n} \in \IZ^8} q^{\half \vec{n}\cdot \vec{n}} ~\Tr (X_1^{n_1} X_2^{n_2} X_3^{n_3} X_2^{n_4} X_1^{n_5} X_2^{n_6} X_3^{n_7} X_2^{n_8}) \\
    &= \frac{1}{(q)_\infty^6} \sum_{\vec{n} \in \IZ^8} q^{\half \vec{n}\cdot \vec{n} - n_2 n_3 + n_5 n_6 - (n_5+n_7)(n_2+n_4+n_6)} ~\delta_{n_1+n_3+n_5+n_7, 0} \cdot \delta_{n_2+n_4+n_6+n_8, 0} \\
    &= \frac{1}{(q)_\infty^6} \sum_{\vec{n} \in \IZ^6} q^{\vec{n}^T K(E_6) \vec{n}} \\ 
\end{split}
\end{align}
where 
\begin{align}
K(E_6) = \left(
    \begin{array}{cccccc}
        2 & 1 & 1 & 1 & 1 & 1 \\
        1 & 2 & 0 & 1 & 0 & 1 \\
        1 & 0 & 2 & 1 & 1 & 1 \\
        1 & 1 & 1 & 2 & 0 & 1 \\
        1 & 0 & 1 & 0 & 2 & 1 \\
        1 & 1 & 1 & 1 & 1 & 2 
    \end{array}
\right) \ . 
\end{align}
Again we find that $K(E_6) = U C(E_6) U^T$ with a  $GL(6,\mathbb{Z})$ matrix
\begin{align}
U=\left(
    \begin{array}{cccccc}
        1 & 0 & 0 & 0 & 0 & 0 \\
        0 & -1 & 0 & 0 & 0 & 0 \\
        0 & -1 & -2 & -1 & 0 & -1 \\
        0 & -1 & -1 & -1 & 0 & -1 \\
        1 & 1 & 1 & 1 & 1 & 1 \\
        0 & -1 & -1 & -1 & 0 & 0 
    \end{array}
\right) \ .
\end{align}
This allows us to write
\begin{align}
\begin{split}
I_2(q) &= \frac{1}{(q)_\infty^6} \sum_{\vec{n} \in \IZ^6} q^{\vec{n}^T C(E_6) \vec{n}} \\
    &= 1 + 78q + 729 q^2 + 4382q^3 + 19917 q^4 + 77274 q^5 + \cdots \ , 
\end{split}
\end{align}
which is the character of the affine $E_6$ algebra at level 1. Once again, the Sugawara central charge of $(E_6)_1$ is $c=6$, which agrees with our expectation. 

\subsection{$(A_1, A_4)$ theory}
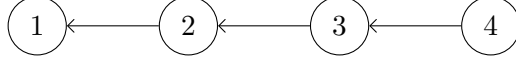
\begin{figure}
  \centering
  \begin{tikzpicture}
			\node[W] (1) at (0,0){1};
		 	\node[W] (2) at (2,0) {2};
            \node[W] (3) at (4,0) {3};
            \node[W] (4) at (6,0) {4};
			\draw[<-] (1)--(2);
            \draw[<-] (2)--(3);
            \draw[<-] (3)--(4);
  \end{tikzpicture}
  \caption{\label{fig:A1A4quiver}
  BPS quiver for the $(A_1, A_4)$ theory in the linear chamber}
\end{figure}

In this section, we consider $(A_1, A_4)$ Argyres-Douglas theory, which is of rank-2. Its BPS quiver (in some particular chamber) is given in figure \ref{fig:A1A4quiver}. 
The monodromy operator at this chamber is given as 
\begin{align}
    M = \Psi_q (X_{\gamma_1}) \Psi_q (X_{\gamma_2}) \Psi_q (X_{\gamma_3}) \Psi_q (X_{\gamma_4}) \Psi_q (X_{-\gamma_1}) \Psi_q (X_{-\gamma_2}) \Psi_q (X_{-\gamma_3}) \Psi_q (X_{-\gamma_4}) \ . 
\end{align}
One can rewrite this operator using \eqref{eq:ids} to get
\begin{align}
    M = \Theta(X_{\gamma_1}) \Psi_q (X_{\gamma_2 - \gamma_1}) \Theta(X_{\gamma_2}) \Psi_q (X_{\gamma_3 - \gamma_2}) \Theta(X_{\gamma_3}) \Psi_q (X_{\gamma_4 - \gamma_3}) \Theta(X_{\gamma_4}) \ . 
\end{align}
The central charge for the VOA is given as 
\begin{align}
    c_{2d}^{(N)} = \frac{40N}{7}-4 \ , 
\end{align}
which gives $c=\frac{12}{7}$ for $N=1$ and $c=\frac{52}{7}$ for $N=2$.

\subsubsection{$\Tr M$: $\mathrm{osp}(1|4)_1$ }
The trace of the monodromy operator can be evaluated straightforwardly
\begin{align}
\begin{split}
    I_1(q) &= (q)_\infty^4 \Tr M \\
    &= \frac{1}{(q)_\infty^4} \sum_{\vec{n} \in \IZ^4} \sum_{{\vec{k} \in \IZ^3_{\ge 0}}} \frac{q^{\frac{1}{2} \vec{n}\cdot\vec{n}}}{(q)_{k_1}(q)_{k_2}(q)_{k_3}} \Tr X_{\gamma_1}^{n_1-k_1} X_{\gamma_2}^{k_1+n_2-k_2} X_{\gamma_3}^{k_2+n_3-k_3} X_{\gamma_4}^{k_3+n_4} \\
    &= \frac{1}{(q)_\infty^4} \sum_{{\vec{k} \in \IZ^3_{\ge 0}}} \frac{q^{\half \vec{k}^T C(A_3) \vec{k}}}{(q)_{k_1}(q)_{k_2}(q)_{k_3}} 
\end{split}
\end{align}
where $C(A_3)$ is the Cartan matrix for the $A_3$ Lie algebra. Upon expanding in $q$-series, we obtain
\begin{align}
 I_1(q) = 1+6q+12q^2+28q^3 + 57q^4 + 108 q^5 + 191 q^6 + \cdots  \ , 
\end{align}
which was conjectured to be the specialized vacuum character of the affine $\mathrm{osp}(1|4)$ algebra at level 1 in \cite{Creutzig:2024ljv}.

\subsubsection{$\Tr M^2$: $(X_1)_1 $}

Using the wall-crossing identities, we can organize $\Tr M^2$ into
\begin{align}
\begin{split} 
 \Tr M^2= \Tr~&\Psi_q(X_{-\gamma_4})\Psi_q(X_{-\gamma_2})\Theta(X_1)\Theta(X_2)\Theta(X_1)\Theta(X_3)\Theta(X_2) \\
&
\cdot\Theta(X_1)\Theta(X_4)\Theta(X_3)\Theta(X_2)\Theta(X_1)
\end{split} 
\end{align}
Evaluating the trace, we find
\begin{align}
\begin{split} 
I_2(q) &= (q)_\infty^4 \Tr M^2 \\
 &=\frac{1}{(q)_\infty^{6}} \sum_{n_1,n_2\geq 0} \sum_{(k_1,\cdots k_{10})\in \mathbb{Z}^{10}} \frac{q^{\frac12(n_1^2+n_2^2+k_1^2+\cdots k_{10}^2)+ k_1k_2-k_5k_6-k_9k_{10}-k_7k_8-k_5k_8-k_5k_{10}}}{(q)_{n_1}(q)_{n_2}} \\
&\qquad \qquad \qquad \qquad \qquad \cdot\delta_{k_1+k_3+k_6+k_{10}, 0} \cdot \delta_{k_4+k_8, 0} \cdot \delta_{n_1-k_7, 0} \cdot\delta_{-n_2+k_2+k_5+k_9, 0}\ ,
\end{split} 
\end{align}
which can also be written as
\be
I_2(q) = (q)_\infty^4 \Tr M^2 =  \frac{1}{(q)_\infty^6}\sum_{n_1,n_7\geq 0}\frac{1}{(q)_{n_1}(q)_{n_7}}\sum_{\{n_2,\cdots n_6,n_8\} \in \mathbb{Z}^6} q^{\frac12 {\bf n}^tK'_8 {\bf n}}\ ,
\ee
where we defined ${\bf n}=(n_1,\cdots, n_8)$ and 
\be
K_8' = \left(\begin{array}{cccccccc} 2 & 0 & 0 & 0 & 0 & 1 & 0 & 0 \\0 & 2 & 0 & 1 & 0 & 1 & -1 & -1 \\ 0 & 0 & 2 & 0 & 1 & 0 & 1 & 1\\ 0 & 1 & 0 & 2 & -1 & 0 & -1 & -1\\ 0 & 0 & 1 & -1 & 2 & 0 & 1 & 1\\ 1 & 1 & 0 & 0 & 0 & 2 & 0 & 0\\ 0 & -1 & 1 & -1 & 1 & 0 & 2 & 1\\ 0 & -1 & 1 & -1 & 1 & 0 & 1 & 2\end{array}\right)\ .
\ee
Again the matrix $K_8'$ defines a rank eight even positive-definite unimodular, which must be isomorphic to the $E_8$ lattice. One can construct a $SL(2,\mathbb{Z})$ matrix $U'$ such that $K_8'= U' C(E_8) {U'}^T$, where
\be
U' = \left(\begin{array}{cccccccc} 1 & 0 & 0 & 0 & 0 & 0 & 0 & 0\\ 0 & 0 & 1 & 2 & 2 & 1 & 0 & 1\\ 0 & 0 & 0 & 0 & -1 & -1 & 0 & -1\\ 0 & 0 & 1 & 1 & 1 & 1 & 0 & 0 \\ 0  & 0 & 0 & 0 & -1 & -1 & 0 & 0\\ 0 & -1 & -1 & 0 & 0 & 0 & 0 & 0 \\ 0 & 0 & 0 & 0 & 0 & 0&1 & 0 \\ 0 & 0 & -1 & -1 & -2 & -1 & 0 & -1 \end{array}\right)\ .
\ee
Therefore, this basis change leaves the two vectors $\alpha_1$ and $\alpha_7$ invariant. This allows us to write the index in terms of $C(E_8)$:
\be\label{X1 fermionic sum}
I_2(q) = (q)_\infty^4 \Tr M^2 =  \frac{1}{(q)_\infty^6}\sum_{n_1,n_7\geq 0}\frac{1}{(q)_{n_1}(q)_{n_7}}\sum_{\{n_2,\cdots n_6,n_8\} \in \mathbb{Z}^6} q^{\frac12 {\bf n}^tC(E_8) {\bf n}}\ .
\ee
Performing the series expansion, we find
\begin{align} \label{X1 character 11}
 I_2(q) = (q)_\infty^4 \Tr M^2 = 1+156q+2236 q^2 + 17056q^3 + \cdots   \ ,
\end{align}
The character in \eqref{X1 character 11} was found as a solution to the third-order modular differential equation in \cite{Das:2022uoe}. The connection to the intermediate algebra $X_1$ \cite{Mkrtchyan:2012es} was observed in \cite{Lee:2024fxa}, claiming that this gives the character of the intermediate vertex algebra called $(X_1)_1$.

\subsection{$(A_1,D_4)$ theory}

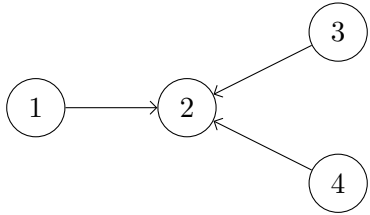
\begin{figure}[b]
  \centering
  \begin{tikzpicture}
			\node[W] (1) at (0,0){1};
		 	\node[W] (2) at (2,0) {2};
            \node[W] (3) at (4,1) {3};
            \node[W] (4) at (4,-1) {4};
			\draw[->] (1)--(2);
            \draw[<-] (2)--(3);
            \draw[<-] (2)--(4);
  \end{tikzpicture}
  \caption{BPS quiver for the $(A_1, D_4)$ theory}
   \label{fig:A1D4quiver}
\end{figure}
The BPS quiver for the $(A_1, D_4)$ theory is given as in the figure \ref{fig:A1D4quiver}. 
The monodromy operator for this theory is given as
\begin{align}
M =  \Psi_q(X_{\gamma_2})\Psi_q(X_{\gamma_4})\Psi_q(X_{\gamma_3})\Psi_q(X_{\gamma_1})\Psi_q(X_{-\gamma_{2}})\Psi_q(X_{-\gamma_{4}})\Psi_q(X_{-\gamma_{3}})\Psi_q(X_{-\gamma_{1}}) \ .
\end{align}
This theory has $SU(3)$ flavor symmetry, which is of rank 2. The flavor charges are given by
\begin{align}
    \gamma_{f_1} = \gamma_4 - \gamma_3 \ , \quad \gamma_{f_2} = \gamma_1 - \gamma_2 \ , 
\end{align}
The central charge for the VOA is given as 
\begin{align}
    c_{2d}^{(N)} = 6 N - 2 \ . 
\end{align}

\subsubsection{$\Tr M$: $(D_4)_1$}

Using the wall-crossing identities, we find
\be
\Tr M = \Theta(X_{\gamma_1})\Theta(X_{\gamma_2})\Theta(X_{\gamma_3})\Theta(X_{\gamma_4})\Theta(X_{\gamma_2})\Theta(X_{\gamma_4})
\ee
Evaluating the trace, we find
\begin{align}
\begin{split}
I_1(q) = (q)_\infty^2 \Tr M &= \frac{1}{(q)_\infty^4} \sum_{(n_1,n_2,n_3,n_4)\in \mathbb{Z}^4} q^{n_1^2 + n_2^2 + n_3^2 + n_3 n_4 + n_4^2 + n_1 (n_3 + n_4) - n_2 (n_3 + n_4)} \\
& = \frac{1}{(q)_\infty^4} \sum_{{\bf n}\in \mathbb{Z}^4} q^{\frac12{\bf n}^t K_4 {\bf n}} \\
&= \frac{1}{(q)_\infty^4} \sum_{{\bf n}\in \mathbb{Z}^4} q^{\frac12{\bf n}^t C(D_4) {\bf n}}
\end{split}
\end{align}
where $C(D_4)$ is the Cartan matrix of $D_4$. The last equality comes from the existence of a $GL(4,\mathbb{Z})$ matrix
\begin{align}
    U = \left( \begin{array}{cccc} 1 & 0 & 0 & 0 \\ 1 & 1 & 0 & -1 \\ 1 & 0 & 1 & 0 \\ 0 & 0 & 0 & -1 \end{array} \right) \ , 
\end{align}
such that $U^tC(D_4)U=K_4$.
This expression coincides with the vacuum character of the VOA $(D_4)_1$, which has the $q$-expansion
\be
I_1(q) = (q)_\infty^2\text{Tr}M = 1+28q + 134q^2 + 568q^3+\cdots\ . 
\ee
Once again, the Sugawara central charge for the $(D_4)_1$ is $c=4$, agreeing with our expectation based on BPS monodromy.

\section{Intermediate vertex algebras from twisted 3d SCFTs} \label{sec:3dSCFT}

The BPS monodromy trace allows us to write the putative character in the form of the Nahm sum (or fermionic expression). This expression can be interpreted as a half-index of certain 3d ${\mathcal{N}=2}$ Chern-Simons matter theory \cite{Gaiotto:2024ioj}. A large class of these 3d theories exhibit supersymmetry enhancement to $\CN=4$ theories along the renormalization group flow, which admit full topological twists that we call the A- and B-twist. If the UV boundary condition flows to an IR boundary condition that is deformable to the A- or B-twist, the half-index can be identified with the boundary chiral algebra of corresponding TFT, which can be identified with the BPS monodromy trace. 

In this section, we explore this idea for two particular examples that are closely related to the so-called intermediate vertex subalgebras $(E_{7\half})_1$ and $(X_1)_1$, which we obtained from $(A_1, A_2)$ and $(A_1, A_4)$ theory respectively from the BPS monodromy operators in the previous section. 
 
\subsection{$``(E_{7\half })_1"$} \label{sec:3dE7half}

\subsubsection{An $\CN=2$ Lagrangian description}

We start from an observation that the Nahm sum formula \eqref{nahm E7.5} can be thought of as a half-index of a simple 3d $\CN=2$ Chern-Simons matter theory,
\be\label{N=2 description E7.5}
\CN=2~~U(1)^8_K ~+~ 1~\text{ chiral multiplet}\ , 
\ee
where the chiral multiplet has charge 1 under the first $U(1)$ factor of the gauge group. The (effective) mixed Chern-Simons level $K$ is identified with the Cartan matrix of $E_8$, $C(E_8)$. This theory can be represented as a quiver diagram depicted in figure \ref{fig:quiverE7.5}. 
\begin{figure}
  \centering
  \begin{tikzpicture}
			\node[W] (1) at (0,0){};
		 	\node[W] (2) at (1.2,0){};
            \node[W] (3) at (2.4,0){};
            \node[W] (4) at (3.6,0){};
            \node[W] (5) at (4.8,0){};
            \node[W] (6) at (6,0){};
            \node[W] (7) at (7.2,0){};
            \node[W] (8) at (4.8,1.2){};
            \node[R] (9) at (0,-1.2){};
			\draw (1)--(2);
            \draw (2)--(3);
            \draw (3)--(4);
            \draw (4)--(5);
            \draw (5)--(6);
            \draw (6)--(7);
            \draw (5)--(8);
            \draw[dashed] (1)--(9);
  \end{tikzpicture}
  \caption{\label{fig:quiverE7.5}
    `Quiver diagram' for the $``E_{7\half}"$ theory. Each solid line corresponds to a mixed CS interaction with level $-1$ (not the bifundamental chiral multiplet), and each of the gauge group factors has the effective CS level $2$. The dashed line corresponds to the chiral multiplet whose charge is $1$. }
\end{figure}
\noindent where each of the solid lines corresponds to a mixed CS interaction with level $-1$, and each of the gauge group factors has the effective CS level $2$. 

The gauge theory description \eqref{N=2 description E7.5} has a global symmetry $U(1)_R\times U(1)_A$, where the factor $U(1)_A$ is the flavor symmetry that corresponds to a particular linear combination of the $U(1)$ topological symmetries of the eight $U(1)$ gauge group factors. 
All the other topological symmetries decouple in the infrared, since there is no gauge-invariant operator charged under such symmetries. It can also be read from the superconformal index we discuss below. 
In the following, we provide evidence that the theory flows to a superconformal theory with $\CN=4$ supersymmetry enhancement.

\subsubsection{Supersymmetry enhancement}

It is convenient to identify the $U(1)_A$ global symmetry with
\be
U(1)_A = U(1)_{\alpha_1} := 2U(1)_1^{\text{top}}-U(1)_2^\text{top}\ , 
\ee
which is a linear combination of the topological symmetries for the first two $U(1)$ gauge group factors. Let us consider a twisted R-symmetry
\be\label{mixing R}
R_\nu = R_0 + \nu A\ ,
\ee
with $\nu\in\mathbb{R}$, where $R_0$ is the reference R-charge which can be chosen to be the superconformal R-charge. We perform F-maximization  \cite{Jafferis:2010un} to fix the superconformal R-charge in the infrared. 

It is useful to compute the superconformal index to test supersymmetry enhancement. The index is defined as 
\begin{align}
    I(q, \eta; \nu) = \Tr (-1)^{R_\nu} q^{\frac{R_\nu}{2}+j_3} \eta^A \ , 
\end{align}
which can be computed via supersymmetric localization \cite{Kim:2009wb, Imamura:2011su}. 
We find that the superconformal index at the IR fixed point has an expansion 
\be
I(q, \eta; \nu=0) = 1-q - \left(\eta + \eta^{-1}\right) q^{3/2} -2q^2 - \left(\eta + \eta^{-1}\right)q^{5/2} - 2q^3 - \cdots\ ,
\ee
where $\eta$ is the $U(1)_A$ fugacity. The existence of the term $-\left(\eta + \eta^{-1}\right) q^{3/2}$ signals the presence of the extra supercurrent multiplet in the $\CN=4$ superconformal algebra \cite{Cordova:2016emh, Evtikhiev:2017heo}.\footnote{There exist two gauge invariant 1/4-BPS dressed monopole operators in the theory, $\phi_1 V_{\pm \theta}$, which can contribute to this term. Here $V_{m}$ is a bare monopole operator with the flux vector $m$, and $\theta=(2,3,4,5,6,4,2,3)$ is the highest root of the $E_8$ algebra.} Furthermore, this expression coincides with the superconformal index of the minimal $\CN=4$ rank-zero theory $\CT_{\text{min}}$, which was first considered in \cite{Gang:2018huc}. This is a strong signal that the $U(1)_K^8$ gauge theory is IR dual to the gauge theory discussed \emph{loc.cit.}, which flows to $\CT_{\textrm{min}}$. 

 The low-energy $\CN=4$ theory admits two inequivalent topological twists, corresponding to the choice $\nu=\pm 1$, which in turn give rise to two distinct TFTs: the A- and B-twists, respectively. The twisted partition function on a circle bundle over a genus-$g$ Riemann surface can be computed \cite{Closset:2016arn, Closset:2017zgf, Closset:2018ghr, Nekrasov:2009uh, Nekrasov:2014xaa} to extract a part of the modular data of the boundary algebra, 
 \be\label{part of modular}
 |S_{0\alpha}| \quad \text{ and } \quad T_{\alpha\alpha}\ , 
 \ee
 up to an overall phase. (See e.g., \cite{Gang:2023rei} for a related discussion.)
For the $\nu=1$ twist, we find that the result is compatible with the following modular representation (up to an overall phase factor that we do not keep track of): 
\be\label{modular E7.5}
S = \frac{2}{\sqrt 5} \left(\begin{array}{cc}\sin (2\pi/5) & \sin (\pi/5) \\ \sin(\pi/5) & - \sin(2\pi/5)\end{array}\right)\ ,\quad T = \text{diag}\left( e^{2\pi i(-19/60)}, e^{2\pi i(29/60)}\right)\ ,
\ee
which reproduces the modular data of $(\hat E_{7 \frac12})_1$ \cite{kawasetsu2014intermediate}. Note that the same data, \eqref{part of modular}, obtained from the bulk theory is also compatible with that of the Lee-Yang minimal model $M(2,5)$, whose modular representation can be written as
\be
S_{LY} = \frac{2}{\sqrt 5} \left(\begin{array}{cc}-\sin (2\pi/5) & \sin (\pi/5) \\ \sin(\pi/5) &  \sin(2\pi/5)\end{array}\right)\ ,\quad T_{LY} = \text{diag}\left( e^{2\pi i(11/60)}, e^{2\pi i(-1/60)}\right)\ .
\ee
These two modular data produce the same partition functions on a large class of Seifert manifolds up to an overall phase, which we do not keep track of in the twisted partition function computation.

Our analysis gives strong evidence that the IR fixed point of the theory \eqref{N=2 description E7.5} is dual to $\CT_{\text{min}}$, possibly up to a multiplication by an invertible topological field theory (TFT).

\subsubsection{Boundary condition}

We now turn to the partition function of the gauge theory on $D^2\times S^1$, which can be identified with the half-index counting the local operators on the $T^2$ boundary. We consider the supersymmetric Dirichlet boundary conditions to all the chiral and vector multiplets of the UV gauge theory, as extensively discussed in \cite{Dimofte:2017tpi}. In order for this boundary condition to be compatible with the topological twist of the IR $\CN=4$ SCFT, it must be deformable to the $\nu=1$ or $\nu=-1$ twist, in the sense introduced in \cite{Costello:2018fnz}. One could attempt to analyze the boundary conditions of the topologically twisted theory via deformations of the holomorphic-topological twist of the $\CN=2$ Lagrangian, as discussed in \cite{Ferrari:2023fez} for simple examples.
However, as illustrated in \emph{loc.cit}, determining whether a given UV boundary condition is deformable is subtle, due to the presence of superpotential terms (involving monopole operators) that must be included in order to pass to the fully topological theory. A detailed analysis of this issue will be addressed in future work; below we simply provide an observation that a formal evaluation of the half-index of the UV gauge theory with supersymmetric Dirichlet boundary conditions gives rise to an interesting modular function which we obtained from the BPS monodromy oprators as discussed in the previous section. With the choice $\nu=1$, we find
\be
I_{\text{half}}(q;s_1,\cdots s_8) = \frac{1}{(q)_\infty^8} \sum_{{\bf n}\in \mathbb{Z}^8}  q^{\frac12 {\bf n}^t C(E_8){\bf n}} (q^{1-n_1}s_1^{-1};q)_\infty\prod_{i,j=1}^8 s_{i}^{C(E_8)_{ij}n_j}\ ,
\ee
where $s_i$'s are fugacity for the boundary currents $U(1)_\partial^8$. Specializing $s_i\rightarrow 1$, we obtain
\be
I_{\text{half}}(q;1,\cdots, 1) = \frac{1}{(q)_\infty^7}\sum_{n_1\geq 0}\sum_{\{n_2,\cdots, n_8\}\in \mathbb{Z}^7} \frac{q^{\frac12 {\bf n}^t C(E_8) {\bf n}}}{(q)_{n_1}}\ ,
\ee
which coincides with the character of the intermediate algebra \eqref{nahm E7.5}.

\subsubsection{Other modules}

We conjecture that the Wilson line $W_1$ of gauge charge $(2,-1,0,\cdots,0)$ flows to the line operator that corresponds to another module whose character forms a vector-valued modular function with \eqref{E7.5 character 11}. Let ${\bf x} = (x_1,\cdots, x_8)$ be the gauge holonomy variables for the UV gauge theory. One can check that the following relation holds
\be
\langle W_{\alpha}\rangle_\beta = L^\alpha({\bf x}^{(\beta)}) =
\frac{S_{\alpha\beta}}{S_{0\beta}}\ ,
\ee
for $L^0=1$ and $L^1=x_1^2x_2^{-1}$,
where $\bf x^{(0)}$ and $\bf x^{(1)}$ are the two solutions to the Bethe equation of the original gauge theory, which is a system of 8 polynomial equations in ${\bf x} = \{x_a= e^{2\pi i u_a}\}$,
\be
P_a({\bf x}) = \exp\left[2\pi i\frac{\partial W(u)}{\partial u_a}\right]=1\ ,
\ee
where $W$ is the Coulomb branch twisted effective superpotential of the gauge theory. See, e.g., \cite{Gang:2023rei} for more details.

The character of this module can be obtained by inserting the simple line $W_1$. Specializing to the limit $s_i\rightarrow 1$, we find
\begin{align}
\begin{split} 
I_{\text{half}}[W_1](q;1,\cdots, 1) &= \frac{1}{(q)_\infty^7}\sum_{n_1\geq 0}\sum_{\{n_2,\cdots, n_8\}\in \mathbb{Z}^7} \frac{q^{\frac12 {\bf n}^t C(E_8) {\bf n}+ 2n_1-n_2}}{(q)_{n_1}}\\
&=\frac{1}{(q)_\infty^7}\sum_{n_1\geq 1}\sum_{\{n_2,\cdots, n_8\}\in \mathbb{Z}^7} \frac{q^{\frac12 {\bf n}^t C(E_8) {\bf n}}}{(q)_{n_1-1}} \ ,\\
&=57+1102q+9367q^2 +57362q^3 +\cdots\ ,
\end{split} 
\end{align}
which is another modular invariant character of $(E_{7\half})_1$.

\subsection{$``(X_1)_1"$}

\subsubsection{An $\CN=2$ Lagrangian description}

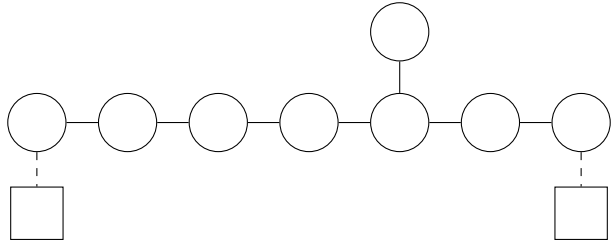
\begin{figure}
  \centering
  \begin{tikzpicture}
			\node[W] (1) at (0,0){};
		 	\node[W] (2) at (1.2,0){};
            \node[W] (3) at (2.4,0){};
            \node[W] (4) at (3.6,0){};
            \node[W] (5) at (4.8,0){};
            \node[W] (6) at (6,0){};
            \node[W] (7) at (7.2,0){};
            \node[W] (8) at (4.8,1.2){};
            \node[R] (9) at (0,-1.2){};
            \node[R] (10) at (7.2, -1.2){};
			\draw (1)--(2);
            \draw (2)--(3);
            \draw (3)--(4);
            \draw (4)--(5);
            \draw (5)--(6);
            \draw (6)--(7);
            \draw (5)--(8);
            \draw[dashed] (1)--(9);
            \draw[dashed] (7)--(10);
  \end{tikzpicture}
  \caption{\label{fig:quiverX1}
    `Quiver diagram' for the $``X_{1}"$ theory. Each solid line corresponds to a mixed CS interaction with level $-1$, and each of the gauge group factors has the effective CS level $2$. The dashed line corresponds to the chiral multiplet whose charge is $1$. }
\end{figure}

The Nahm sum type formula for \eqref{X1 fermionic sum} suggests a 3d $\CN=2$ Chern-Simons matter theory description
\be
\CN=2~~U(1)^8_K ~+~ 2~\text{chiral multiplets}\ , 
\ee
where the chiral multiplets has charge 1 under the first $U(1)$ factor and the seventh $U(1)$ factor of the gauge group. The (effective) mixed Chern-Simons level $K$ is identified with the Cartan matrix of $E_8$, $C(E_8)$. This theory can be represented as a quiver diagram depicted in figure \ref{fig:quiverX1}. 
This theory is almost identical to the one discussed in section \ref{sec:3dE7half}, except that we added one extra charged chiral multiplet. 

The gauge theory description allows a half-BPS monopole operator of the form:
\be
V_{(2, 2, 2, 2, 2, 1, 0, 1)}\phi_7\ ,
\ee
where $V_{\bf m}$ is a bare monopole operator with gauge flux ${\bf m}$ and $\phi_7$ is the scalar field in the second chiral multiplet (attached to the seventh node). Deforming the theory with the monopole superpotential, the global symmetry that acts non-trivially on the IR theory is $U(1)_A\times U(1)_R$, as in the previous example. 

\subsubsection{Supersymmetry enhancement}

We identify the $U(1)_A$ symmetry as
\be
U(1)_A = U(1)_1^{\text{top}} - U(1)_2^{\text{top}} + U(1)_7^{\text{top}}\ .
\ee
The superconformal index at the IR fixed point can be computed to give 
\be
I(q, \eta; \nu=0) = 1-q - \left(\eta + \eta^{-1}\right) q^{3/2} -2q^2 -\eta q^{5/2}+\left(-1+\frac{1}{\eta^2}\right) q^3+ \cdots\ ,
\ee
where again $\eta$ is the $U(1)_A$ fugacity. One can check that the first several coefficients of the expansion coincide with the superconformal index of the $\CN=4$ rank-zero SCFT called $\CT_{2}$, which was considered first in \cite{Gang:2023rei}. This suggests that our theory is IR dual to the $\CT_{2}$ theory.  

The low-energy $\CN=4$ superconformal theory can be topologically twisted to produce two distinct TFTs, which correspond to the choice $\nu=\pm 1$, where $\nu$ is a twisting parameter we introduced in  \eqref{mixing R}. For the $\nu=1$ twist, the modular data \eqref{part of modular} extracted from the partition function on Seifert manifolds are compatible with the following representation:
\be
S =\frac{2 \sin(\pi/7)}{\sqrt{7}}\begin{pmatrix} d^2 \!-\! 1 & d & 1\\ d & -1 & 1\! -\! d^2\\ 1 & 1 \!-\! d^2 & d \end{pmatrix}\ ,
\ee
where $d=2\cos(\pi/7)$\ , and
\be
\quad T = e^{2\pi i (-13/42)}\text{diag}\left(1, e^{2\pi i(6/7)}, e^{2\pi i (4/7)} \right)\ .
\ee
We can check that this modular data is almost, but not quite identical to, that of the Virasoro minimal model $M(2, 7)$, which can be obtained via topologically twisted $\CT_2$. Each component of modular data differs by a phase. 
Therefore, we claim that this theory is dual to $\CT_2$ up to a possible invertible topological field theory.

\subsubsection{Boundary condition}

Again for this example, we formally evaluate the half-index of the gauge theory with the supersymmetric Dirichlet boundary conditions for all the $\CN=2$ multiplets. For $\n=1$, we find
\be
I_{\text{half}}(q;s_1,\cdots s_8) = \frac{1}{(q)_\infty^8} \sum_{{\bf n}\in \mathbb{Z}^8}  q^{\frac12 {\bf n}^t C(E_8){\bf n}} (q^{1-n_1}s_1^{-1};q)_\infty(q^{1-n_7}s_7^{-1};q)_\infty\prod_{i,j=1}^8 s_{i}^{C(E_8)_{ij}n_j}\ ,
\ee
where again $s_i$'s are the fugacities for boundary $U(1)^8_\partial$ currents. Specializing to $s_i\rightarrow 1$, we recover the vacuum character of an intermediate vertex algebra $(X_1)_1$  \cite{Mkrtchyan:2012es}:
\be
I_{\text{half}}(q;1,\cdots, 1)=\sum_{n_1,n_7\geq 0}\frac{1}{(q)_{n_1}(q)_{n_7}}\sum_{\{n_2,\cdots n_6,n_8\} \in \mathbb{Z}^6} q^{\frac12 {\bf n}^tC(E_8) {\bf n}}\ .
\ee

\subsubsection{Other modules}

We conjecture that the following two Wilson lines
\be\label{wilson lines X1}
W_1 = (1,-1,0,\cdots,0,1,0)\ ,\quad W_2 = (0,-1,0,\cdots,0 ,1)\ ,
\ee
flow to the line operators, which correspond to the two modules of the boundary algebra whose characters transform as a vector-valued modular form together with \eqref{X1 character 11}. 
Let ${\bf x} = (x_1,\cdots, x_8)$ be the gauge holonomy variables for the UV gauge theory. One can check that the following relation holds
\be
\langle W_{\alpha}\rangle_\beta = L^\alpha({\bf x}^{(\beta)}) =
\frac{S_{\alpha\beta}}{S_{0\beta}}\ ,
\ee
for $L^0=1$, $L^1=x_1^1x_2^{-1}x_7$, and $L^2= x_2^{-1}x_8$,
where $\{{\bf x^{(0)}}\}$, $\{{\bf x^{(1)}}\}$, and $\{{\bf x^{(2)}}\}$ are the three solutions to the Bethe equation of the gauge theory.

One can compute the half-indices with an insertion of these line operators \eqref{wilson lines X1}, which give
\begin{align}
\begin{split} 
I_{\text{half}}[W_1](q;1,\cdots, 1)&=\sum_{n_1,n_7\geq 0}\frac{1}{(q)_{n_1}(q)_{n_7}}\sum_{\{n_2,\cdots n_6,n_8\} \in \mathbb{Z}^6} q^{\frac12 {\bf n}^tC(E_8) {\bf n}}q^{n_1-n_2+n_7}\\
&=13+364q + 3302q^2 +\cdots\ ,
\end{split}
\end{align} 
and
\begin{align}
\begin{split} 
I_{\text{half}}[W_2](q;1,\cdots, 1)&=\sum_{n_1,n_7\geq 0}\frac{1}{(q)_{n_1}(q)_{n_7}}\sum_{\{n_2,\cdots n_6,n_8\} \in \mathbb{Z}^6} q^{\frac12 {\bf n}^tC(E_8) {\bf n}}q^{-n_2+n_8}\\
&=78+1288q + 10465q^2 +\cdots\ ,
\end{split}
\end{align} 
which coincide with the $q$-expansion of $\chi_{4/7}$ and $\chi_{6/7}$ of $(X_1)_1$ discussed in \cite{Lee:2024fxa}. They appear as solutions to the third-order modular linear differential equation \cite{Das:2022uoe}.


\section{Discussion} \label{sec:disc}

In this paper, we have discovered a correspondence between multiple powers of the BPS monodromy operator for a given 4d $\CN=2$ SCFT and a family of vertex algebras. This connection can be understood via $U(1)_r$-twisted compactification of the 4d theory, which gives rise to a 3d $\CN=4$ SCFT. Upon suitable topological twisting of the 3d theory and by putting the theory on a space with a boundary, we obtain the corresponding boundary algebra. The family of vertex algebras arises from the multiplicity of twisting as we reduce the 4d theory to 3d.

We have found various vertex algebras realized from the BPS monodromy of Argyres-Douglas theories, especially a series of vertex algebras given by affine Lie algebras in the Deligne-Cvitanovi\'c exceptional series of level 1. Remarkably, they include the characters of the `intermediate vertex algebras' such as $(E_{7\half})_1$ and $(X_1)_1$. Note that these characters we obtain also appear in a twisted sector of rational vertex operator algebras. For example, $(E_{7\half})_1$ characters can be obtained from the rational and $C_2$-cofinite W-algebra $W_{-5}(E_8, f_\theta)$ \cite{kawasetsu2018algebras}. 

These formulas motivate us to conjecture the existence of certain $\CN=2$ abelian Chern-Simons matter theories that flow in the infrared to $\CN=4$ SCFTs with zero-dimensional Coulomb and Higgs branch. The 3d $\CN=4$ theories associated with $(E_{7\half})_1$ and $(X_1)_1$ display another interesting feature: their superconformal indices and the twisted partition functions (on a circle fibration over a Riemann surface) agree (up to an overall phase) with that of minimal $\CN=4$ SCFTs $\CT_{\textrm{min}} = \CT_1$ and $\CT_2$ studied in \cite{Gang:2018huc, Gang:2021hrd} whose boundary VOAs are Virasoro minimal models. This strongly suggests that our gauge theories are IR dual to $\CT_1$ and $\CT_2$, possibly up to an invertible TQFT. 

The UV boundary condition that directly yields the Nahm sum expressions for the characters is the supersymmetric Dirichlet boundary conditions discussed in \cite{Dimofte:2017tpi}. While this boundary condition is always valid in the $\CN=2$ theory preserving 2d $\CN=(0,2)$ algebra, in order for it to be compatible with the full topological twist in the IR $\CN=4$ theory, the boundary condition must be deformable to the twist in the bulk. In this work, we defer a detailed analysis of the deformability of the boundary condition to future study, and instead formally evaluate the Dirichlet half-index, observing that it gives rise to the modular-invariant character of the intermediate algebras $(E_{7\half})_1$ and $(X_1)_1$. It would be interesting to explicitly compute the boundary OPEs of these 3d topological field theories to clarify their precise relations to the corresponding algebras. We plan to return to this question in future work. 

In some sense, our construction provides a generalization of the SCFT/VOA correspondence of \cite{Beem:2013sza}, which comes from $\Tr M^{-1}$ in our language. However, unlike the case of \cite{Beem:2013sza}, where the associated VOA can be defined purely in terms of 4d superconformal field theory data by passing to certain cohomology, we do not have such a direct construction. It would be interesting to clarify the connection between other vertex algebras coming from $\Tr M^N$ and 4d physics at the conformal point. 

There are numerous questions we can ask regarding our connection between 4d $\CN=2$ SCFT, 3d $\CN=4$ SCFT, and VOAs. We would like to conclude by making some comments on future directions. 
\begin{itemize}
    \item Connection to the holomorphic modular bootstrap program \cite{Mathur:1988gt, Mathur:1988na}: We find that for a given AD theory, we get a family of vertex algebras whose characters solve the MLDEs of the same degree \cite{arakawa2018quasi, mason2018vertex, Das:2022uoe}. Why is this true? 
    It is conjectured that for any 4d $\CN=2$ SCFT, the Schur index should obey a finite order MLDE \cite{Beem:2017ooy}. Is there any connection between the two? 
    
    \item Other powers of monodromy operator: For Argyres-Douglas theories, there exists a positive integer $n$, such that $n$-th wrapping of the Janus configuration trivializes the $U(1)_r$-twisting \cite{Cecotti:2010fi}. Therefore, it is natural to expect that some sort of periodicity exists in the higher powers of monodromy traces. However, it appears that $\Tr M^n$ does not give rise to a converging $q$-series. It would be interesting to see if there is a way to systematically regularize this sum as was done in \cite{Cordova:2016uwk}.

    \item Fusions and modular properties: Characters for other simple modules can be obtained via inserting surface defects \cite{Cordova:2017ohl, Cordova:2017mhb}. The trace formula for the $S^3$-partition functions and the modular data can be obtained for the higher powers of monodromy as well \cite{Gaiotto:2024ioj}, which will be discussed in an upcoming work \cite{Go:2025ixu}. 

    \item Higher-rank AD theories: It would be interesting to further investigate the correspondence to higher-rank Argyres-Douglas theories. For the $(A_1, G)$ type AD theories, the BPS quiver is known, and we can straightforwardly compute the monodromy traces. For other $(G, G')$ theories, the corresponding BPS quiver does not encode the full BPS spectrum. Therefore, it is not known how to compute the monodromy operator. Instead, one can use $\CN=1$ Lagrangian description \cite{Agarwal:2017roi, Benvenuti:2017bpg} for certain cases of AD theories and follow the strategy of \cite{ArabiArdehali:2024vli, ArabiArdehali:2024ysy}. However, this method becomes computationally demanding for higher powers of monodromy (or higher sheets) and higher rank. 

    \item Non-Argyres-Douglas theories: In our study, it was crucial to have a Coulomb branch operator with fractional $U(1)_r$ charge to have a non-trivial twisting. However, nothing stops us from computing the higher powers of the monodromy operator for arbitrary $\CN=2$ SCFTs. It would be interesting to ask if it gives us any interesting vertex algebras. 
    
    \item VA-valued TQFTs from class $\CS$: It is known that one can construct a vertex operator algebra-valued TQFT \cite{Gadde:2009kb, Beem:2014rza} by considering class $\CS$ theories \cite{Gaiotto:2009we, Gaiotto:2009hg}. Here VOA is associated with $\Tr M^{-1}$ of the corresponding class $\CS$ theory. It is interesting to ask if there can be a TQFT structure for the VAs constructed from $\Tr M^N$ of class $\CS$ theory. The class $\CS$ also included a large set of Argyres-Douglas theories \cite{Xie:2012hs, Wang:2015mra}, whose associated VOA is particularly simple \cite{Xie:2016evu, Song:2017oew}. It would be interesting to ask if such a structure persists.  

\end{itemize}

\begin{acknowledgments}

We thank T. Arakawa, D. Gaiotto, and D. Gang for useful comments and discussions. The work of HK is supported by the National Research Foundation of Korea (NRF) grant NRF-2023R1A2C1004965. 
The work of JS is supported by the National Research Foundation of Korea (NRF) grant RS-2023-00208602 and also by the POSCO Science Fellowship of POSCO TJ Park Foundation.
This work is also supported by the National Research Foundation of Korea (NRF) grant RS-2024-00405629.

\end{acknowledgments}

\bibliographystyle{jhep}
\bibliography{voabps}

\end{document}